# Characteristics and Motivations of Players with Disabilities in Digital Games

**Work in Progress**

Jen Beeston[1][0000-0002-9284-2867], Christopher Power[1,2][0000-0001-9486-8043], Paul Cairns [1,2][0000-0002-6508-372X], and Mark Barlet[2][0000-0001-7081-0314]

[1] Department of Computer Science Deramore Lane, University of York, Heslington, York, YO10 5GH, UK
[2] The AbleGamers Charity, PO Box 508, Charles Town WV 25414 USA
`{jen.beeston, christopher.power, paul.cairns}@york.ac.uk,`
`mark@ablegamers.org`

**Abstract.** In research and practice into the accessibility of digital games, much of the work has focused on how to make games accessible to people with disabilities. With an increasing number of people with disabilities playing mainstream commercial games, it is important that we understand who they are and how they play in order to take a more user-centered approach as this field grows. We conducted a demographic survey of 230 players with disabilities and found that they play mainstream digital games using a variety of assistive technologies, use accessibility options such as key remapping and subtitles, and they identify themselves as gamers who play digital games as their primary hobby. This gives us a richer picture of players with disabilities and indicates that there are opportunities to begin to look at accessible player experiences (APX) in games.

## 1    Introduction and Background

Currently, video games represent a significant part of our everyday modern lives, with UKIE estimating that between 2.2 and 2.6 billion people play digital games worldwide.  From online activity in social media, Twitch and player communities, we know that players with disabilities are increasingly part of what is a dynamic and growing community of digital game players, however there is very sparse research into who they are as players, the types of games they play and the prevalence of use of assistive technologies and accessibility settings in games.  We surveyed 230 players with disabilities collected as part of the AbleGamers Player Panels programme, to direct future research as to the diversity of this distinctive population of players, and to inform design in terms of the diversity of breadth of technologies that are currently being used in digital games.

Researchers and designers alike acknowledge that there are not only barriers to playing digital games, but also accessibility concerns within the games themselves for those with individual and complex needs. Previously, researchers have considered the barriers that players with disabilities encounter in games, with their efforts focused mainly on how technology can be adapted to enable them to play [1] and on creating bespoke games to investigate how games can be made playable for players with varying disabilities [12,18]. In research focused on games for supporting therapeutic applications, it is acknowledged that because there is such complexity in the symptoms of disabling conditions, it is difficult to know which games to provide or suggest for these potential players. Therefore, research has been directed at tools to help match mainstream games with players for therapy [22]. Otherwise, several charity organizations (e.g. AbleGamers, Gamers Outreach, Special Effect) and advocates of disabled gaming have established community and support for players with disabilities and created information to guide game developers to make adaptations and improvements [3]. Following on from this work, the successful integration of accessibility into many commercial titles means there are many people with disabilities playing online amidst non-disabled players. However, little is known about this audience and their player experiences, and how and why they choose the games they do. It is currently unclear what, if any, technology and accessibility options are being used by players with disabilities. Some academics have found that various methods that can be used to enable play, such as controllers [5], skill assistance [8] and difficulty adjustments [2] may impact upon the experience of play for those using them and for other players when these features are being used. It can be argued that gaming is an inherently social hobby [26,24]. Therefore, for players with disabilities who may use adaptations and non-standard peripherals to play, it is important to consider what effect this has, not only on their experiences in games, but also the social elements of the gaming hobby.

Digital games are widely recognized as a popular, enjoyable and even beneficial activity from which players can derive a sense of wellbeing [13]. Therefore, it is important that access to games should be universal and should not exclude people with disabilities. Most players view games as an end in themselves that provide experiences that are intrinsically valued [15]. However, rather than thinking about how players with disabilities play mainstream games that everyone plays, games are often positioned as a means to an end, particularly for rehabilitation and research is often focused on creating bespoke games [12] or adapting and creating novel controllers [18]. Regardless of whether these are the reasons why these people play games or not, this approach neglects the evidence that there are growing numbers of players with disabilities playing mainstream games alongside everyone else. Digital games are supporting real inclusion but little is known about how players are gaining access to games and the experiences they have when they are playing, whether valued or not.

If we are to move research and practice beyond questions of basic access and enablement [21] it is important that we know more about players in the game space so we can begin to understand their accessible player experiences (APX). Porter and Kientz [20] provide a useful starting point with a survey of 55 players with disabilities collected age, gender, impairment class, platforms played on, and types of games played and was also supplemented by interviews. They found that their participants had some incompatibilities with technologies that were barriers to their gaming, and that their sample tended towards single player games and less towards multiplayer games. Additionally, they spoke to games industry professionals about their current practises in making games accessible for players with disabilities. Their findings suggest that the games industry focus on the things that they are immediately aware of, such as a colleague having a specific need, like colour remapping or subtitles. The motivation of the work presented is to provide a more comprehensive understanding of players with disabilities to inform and extend the focus of subsequent research and practice into accessible games.

## 2   Method

The AbleGamers Player Panels programme was created in a collaboration between the University of York and the AbleGamers Charity to provide a systematic means by which players with disabilities can have a voice in digital games research and development. The aim of the programme is to facilitate organisations looking to do user research or games testing with players with disabilities by putting them in touch with suitable players who have already declared an interest in taking part in such research. To enable such matchmaking and also to provide a better understanding of players with disabilities, the aim of this study is the collection of the necessary demographic information about the players.

The demographic survey was iteratively developed with collaboration between the AbleGamers team and players with disabilities. Players originally registered interest in the Player Panels through the AbleGamers Charity website, where they provided a small amount of demographic information including: contact information, age, current gaming platforms used, game genres played, and their motivation to register. The AbleGamers Player Panels registration requested that players only register if they want to be involved and be contacted by researchers and developers, have access to the internet and could fill out the online survey. The demographic questionnaire was available for all ages and for those with any form of disability that did not prevent them from completing the online questionnaire. This work and further research only included participants over the age of 18 and excluded those who have indicated that they have a cognitive disability. This was to safeguard those for whom consent could not be guaranteed at this stage.

A sample of 7 respondents from the California area took part in a telephone interview to trial further demographic questions and to inform questions about their gaming habits. From this, an initial draft demographic questionnaire was created and feedback elicited from 5 further registered respondents and from AbleGamers staff.

The final demographic survey contained demographic information, such as their contact information, preferences, gaming needs, current habits and technology usage, which can be used to identify participants for opportunities with organisations. Further, participants gave consent for the information to be used by AbleGamers and their partners for purposes of research.

## 3 Results

### 3.1 About the Players

Out of 230 respondents, 156 people identified as male, 52 female, 16 non-binary and 6 preferred not to say. The average age of respondents was 31. When asked about the length of a typical play session for them, 116 respondents said they played between 2 to 4 hours at a time; 48 played 1 to 2 hours, 51 played 5 hours or more, and only 12 people reported a typical session as being 1 hour or less. Respondents were asked to select as many of the items in Table 1 to describe their disabilities as required. It is worth noting that 'Other needs and preferences' was an open text item. This mainly seems to have been used to provide a more detailed description or the medical terms for their disability. To retain confidentiality, this information is not provided here.

**Table 1.** Disability information

| Disability | Respondents |
|---|---|
| Autism | 19 |
| Hard of hearing | 28 |
| Deaf | 8 |
| Upper limb physical disabilities | 128 |
| Lower limb physical disabilities | 120 |
| Mental Health Difficulties | 55 |
| Learning Disabilities (e.g. dyslexia, SLP, ADHD, language etc.) | 29 |
| Blind | 17 |
| Colour vision deficient (e.g. red-green colour blind) | 9 |
| Low vision | 29 |
| Other needs and preferences | 59 |

Many of the respondents identified themselves as gamers (150) and consider it to be their primary hobby (138). There were an almost equal number of people who considered themselves to be hardcore gamers (101) as those who identified as casual gamers (68). Very few people did not consider themselves a gamer (24) or only played games when they have nothing else to do (19).

### 3.2 What Are They Playing?

The gaming platforms rated as being used 'very often' by respondents were PC (152), followed by PlayStation (83), phone (72), Xbox (44), Nintendo Switch (33), and tablet (33). Out of those platforms, Nintendo Switch scored highest in the 'do not play' category, followed by Xbox, tablet, PlayStation, Phone, and then PC. It is quite possible that Nintendo Switch was least played since it was the newest gaming console listed in the options. The game types selected as played most often were Single Player (195), followed by Online Multiplayer (114), Cooperative Multiplayer (71), Competi-tive Multiplayer (63), One vs. One Multiplayer (47), and then Local multiplayer (31). Respondents were asked to provide their top 3 current favourite games. There were 329 different titles provided. Where games received more than one entry, a top favourite games list was created to show which were the most popular games.

**Table 2.** Top favourite games

| Rank | Top favourite games | Respondents | Best-selling games of 2017 by NPD Group |
|---|---|---|---|
| 1 | Destiny 2 | 17 | Call of Duty: WWII |
| 2 | World of Warcraft | 15 | Star Wars: Battlefront II |
| 3 | Overwatch | 14 | Super Mario Odyssey |
| 4 | PlayerUnknown's Battlegrounds | 10 | NBA 2K18 |
| 5 | The Elder Scrolls V: Skyrim | 10 | Mario Kart 8 |
| 6 | The Legend of Zelda: Breath of the Wild | 9 | Madden NFL 18 |
| 7 | Grand Theft Auto V | 9 | PlayerUnknown's Battlegrounds |
| 8 | Super Mario Odyssey | 9 | Assassin's Creed: Origins |
| 9 | Rocket League | 9 | The Legend of Zelda: Breath of the Wild |
| 10 | Stardew Valley | 8 | Grand Theft Auto V |

**Source of ranked list:** https://venturebeat.com/2018/01/18/december-npd-2017-nintendo-switch-leads-the-hardware-pack-in-a-3-29-billion-month/
**Note:** Please see footnotes 2 in section 4 for further top games lists by platform

### 3.3 How Are the Players Accessing Games?

Participants were asked to indicate whether they used any items from a selection of assistive technologies (hardware) and accessibility options (software) or could specify in separate textbox if they used something not listed. Of the assistive technologies, 24 respondents provided information in the 'other' box. Items such as on-screen keyboard and using a converter to use keyboard and mouse on console were mentioned. One respondent mentioned that they used a handheld magnifying glass, but they did not specify exactly what they used this for. Customized controllers or alternative PC mouse were also selected as often used assistive technologies. Popular accessibility options items used were subtitles (108 participants) and key remapping/bindings used by 117 respondents.

Of the 230 participants, 156 people indicated that they did not use any assistive technology, 77 people did not use any accessibility features, and 47 people did not use any assistive technologies or accessibilities features to play with.

**Table 3.** Assistive gaming technology and in-game accessibility options

| Assistive Technology | Respondents | Accessibility options | Respondents |
|---|---|---|---|
| Eye gaze tracking | 5 | Text to speech | 14 |
| Customized controller | 26 | Speech to text | 17 |
| One handed controller | 5 | Subtitles | 108 |
| Screen reader | 17 | Colour blind options | 12 |
| Alternative PC mouse | 25 | Contrast or colour changes | 34 |
| VR headset | 3 | Mouse cursor enlargement | 24 |
| Alternative controller | 10 | Text enlargement | 46 |
| Other technology | 33 | Auditory or screen alerts | 28 |
| | | Key remapping | 117 |
| | | Other option | 28 |

### 3.4 Are They Playing Alone?

Only 3 people indicated that they did not play single player games, therefore the majority of participants play alone at some stage. To ascertain what portion of participants played with others, 5 items were provided to indicate what kinds of multiplayer (MP) game were played; Local, Online, Cooperative, Competitive, One vs. One. This gives some indication of preference; however, it is worth noting that many multiplayer games are a combination of these categories. The most useful items for comparison here are, single player, local multiplayer (co-located play) and online multiplayer. Only 33 participants indicated that they did not play any online multiplayer games compared to 82 who said they did not play local (co-located) multiplayer.

Table 4. What types of multiplayer games are being played

| Game type | Do not play | Sometimes | Very often |
|---|---|---|---|
| Single player | 3 | 32 | 195 |
| Local multiplayer | 82 | 117 | 31 |
| Online multiplayer | 33 | 83 | 114 |
| Cooperative multiplayer (team vs. game) | 56 | 103 | 71 |
| Competitive multiplayer (team vs. team) | 70 | 97 | 63 |
| One vs. one multiplayer | 102 | 81 | 47 |

### 3.5 Who Are They Playing With?

Understandably, local multiplayer games were mostly played with real life friends of participants, though some people played these games with online friends and strangers. This could perhaps indicate that these games might be played in a public setting, such as a gaming centre or an arcade. Online multiplayer games were mostly played with online friends, and almost equally with real life friends and strangers. Of 197 participants that play online multiplayer games, 11 people said that they played them only with real life friends. Similarly, 18 people played co-op MP games (of 174 participants) with only real-life friends, 14 people competitive MP games (of 160), and 20 people one versus one (of 128).

**Table 5.** Who are people playing with in each game type?

| Game type | Real-life friends | Friends of friends | Online friends | Guild of clan members | Strangers |
|---|---|---|---|---|---|
| Local multiplayer | 153 | 44 | 28 | 20 | 18 |
| Online multiplayer | 133 | 84 | 159 | 81 | 131 |
| Cooperative multiplayer (team vs. game) | 124 | 76 | 130 | 62 | 79 |
| Competitive multiplayer (team vs. team) | 110 | 68 | 114 | 55 | 101 |
| One vs. one multiplayer | 88 | 43 | 86 | 33 | 80 |

### 3.6 How Are They Communicating in Play?

The participants who played any form of multiplayer game were asked to specify which communication platforms they used during play very often, sometimes and not at all. The most popular communication platform was Discord (59) followed by: the games own provided chat/voice comms (58), PlayStation Network Chat (33), Xbox Party Chat (26), Skype (25), TeamSpeak (11), Mumble (2), and Ventrilo (1). All of these communication platforms were most often used on PC with the exceptions of PlayStation Network Chat which was mostly used on by those who used PlayStation 'very often' and Xbox Party Chat, used by Xbox players. Of the participants, 24 said that they didn't use any of those communication methods listed. Since we did not offer an open text entry on this item, it is not clear whether they simply did not use communication in game or whether they used some other platform, such as Facebook or WhatsApp.

### 3.7 What Are the Players Reasons for Gaming?

The most popular reason provided for why participants play games was to have fun. All but 3 participants selected this item. Relaxation, challenge, community, and escapism were also commonly selected reasons for play. Interestingly, health based reasons - stress and mental health management were more common than competition for this sample of players. Approximately a third of participants indicated that one of their reasons for playing games was related to pain management.

**Table 6.** Reasons for playing games.

| Why Play? | Respondents |
|---|---|
| To have fun | 227 |
| To help me relax | 199 |
| To challenge myself | 171 |
| To be part of a community | 171 |
| To escape reality | 160 |
| To socialise | 150 |
| To aid in my stress management | 148 |
| To aid with my mental health | 115 |
| To compete with others | 84 |
| To aid in my pain management | 71 |
| Other reason | 22 |

## 4  Discussion

The results show that this sample of players with disabilities are choosing to play mainstream, commercial games. Many of their favourite current games are aligned with current, top/most played games across the common gaming platforms which strongly suggests that the gaming preferences of these players is no different from non-disabled digital game players. While this aligns with Porter and Kientz [20] and Flynn and Lange [10] regarding the desire of people to play mainstream AAA titles, we differ in that our sample shows that more than half of our players favourite games are multiplayer games. Whether this is due to sampling bias, or due to a shift in demographics since that previous work, we have compelling evidence that players are engaging in both single player games, and online, community-based play.

Our findings show that there are some adaptations that are commonly used among this sample, such as customised controllers/PC mouse, subtitles and key remapping. This suggests that even such relatively straight forward adaptations provided in games can help to enable play for many people. PC was the most used gaming platform by participants, which is consistent with common wisdom that up until recently PC gaming was more accessible than consoles as accessibility is more mature on that platform [14]. It will be important to revisit this in the near future now that a number of consoles are integrating middleware solutions for accessibility. Phone was the second most used platform by respondents which may be due to the ubiquity of the smartphone in modern life which is something that people are likely to own anyway rather than a separate platform for gaming.

Many of these players consider themselves to be gamers, and a substantial portion say that they are hardcore gamers which suggests that they identify deeply with the gaming hobby and invest substantial time and effort on the hobby [6]. If this is the case, there are social aspects to consider for these players within gaming, too. Many of the participants indicated that they play a range of different forms of multiplayer games and as such, are gaming with others at least some of the time. Since very few people indicated that they did not play any online multiplayer games, this supports that gaming is a social hobby for players with disabilities much like non-disabled players.

The results suggest that there is a preference for cooperative multiplayer over competitive multiplayer games, though only minor. This could indicate that this sample of players are less inclined to play for competitive reasons, this appears to be supported by the results of participants reasons for gaming. Competition did not appear to be one of the main motivations for this sample to play games.

When participants were required to specify who they played with from a number of options, those who played online multiplayer games play with both real-life friends and strangers about equally, but primarily with online friends. It is not clear whether this is due to formation of online friends through gaming for gaming purposes or simply having fewer real life friendships that extend into gaming. Xu and colleagues [26] consider a 'game as a medium of social relationships' in their work on social relationships in FPS games and, in particular, within the game Halo 3. They conclude that although players had a significant number of real life friends in their 'friend list', players often 'friend' other players that they have met online through games. And that this friendship occurred as a result of playing successfully together multiple times and communicating during the game. Xu et al, [26] also suggest that additional friendships form to include friends of friends (triadic closure [16]) through the game by either player introducing their friend to work together in the game. This may go some way to explain the number of online friendships indicated by our sample.

The results also show that, when playing cooperative games, people played with a higher number of real-life friends than strangers. However, when playing competitive games there was little difference between the numbers of real-life friends and strangers. Additionally, people who played cooperative games, overall, played with less strangers than those who played competitive. This is worthy of deeper exploration to determine whether this trend is linked to their gaming goals, preferences and abilities or rather more influenced by the types of games that they are choosing to play. It is not possible at this stage to determine which games players have in mind when thinking about cooperative (team vs. game) and competitive (team vs. team). The type of game and the way that individual teams are formed within the game may have some influence on whether players are more or less likely to play with strangers. For example, a game like Overwatch (team vs. team) may allow for more random formulation of teams due to its matchmaking system related to player skill and thus, mean that players end up playing more with strangers. Whereas, when playing a game such as World of Warcraft (team vs. game, though not always), teams may be formed over time between players who meet and bond and form friendships [19].

### 4.1 Communication.

Among players that use voice chat communication within games, the communication platform used seems sensibly linked to the gaming platform that they play on. For example, people playing on PlayStation mostly used PlayStation Network Chat. Interestingly, on the most popular gaming platform - PC, players used Discord to chat more so than the voice chat that is provided by individual games. Freeman and Wohn [11], in their study of E-sports players, also found that platforms such as Discord and TeamSpeak were preferred for social interactions with co-players. There could be a number of possible reasons for these findings. Discord is a communication platform for gaming and it has overlay functionality so that it can be used in most popular PC games. It allows users to create specific channels which any person with the channel link can join. Within a channel, smaller subchannels can be created for specifically voice communication and chat. It arguably creates an optimal platform for the creation and maintenance of gaming communities, whereas a games own communication platform may be temporary and limited to individual play sessions. Another possible reason why players may be choosing a third-party voice communication platform is that it allows users to more easily control who they voice chat with. Wadley et al. [25] worked on a grounded theory towards understanding the use of voice chat in online play. They found that voice chat was not always preferred over text-based chat and that 'griefing' among players was felt to be worse in voice chat. They suggest that, although voice chat can be positive for players to build on social connections with other players, it may 'interfere with pseudonymity'. This is because voice chat allows other players to learn things about the speaker through their style of communication and things like nationality or location that could be guessed through the sound of their voice. A reduction in pseudonymity could potentially be an important factor for players with disabilities to allow them to avoid any potential discrimination that they may feel they could face in social play. Further, a platform such as Discord allows players to establish and maintain online friendships, and speak to others during play for strategic communication but also may act as a buffer against 'aggressive' communication that could come through other in-game voice channels.

### 4.2 Reasons for play

The primary reason selected by participants as to why they play games was to have fun. This suggests that participants are motivated to play for the sake of enjoyment and leisure, much like non-disabled players. Personal challenge, being part of a community, and escaping reality were also commonly selected reasons. This is not dissimilar to the findings of Sherry et al. [23] who found that challenge was a main reason given for play. They also found that competition was a significant motivator to play, however, our findings do not match this. Respondents indicated that they played for health reasons, namely stress management, and as an aid to mental health over competition. It is not clear whether this is related to their disabilities or not. However, where some participants have offered other reasons in the text entry, a common theme within these comments was that people were playing for therapeutic reasons. Examples include:

"To help maintain mental sharpness and clarity"
"Physical therapy for hands"
"Combat depression"
"To slowly work through issues/empathy"

This supports that playing games is not only a means to an end for these players, but beneficial for other health-related reasons and of their own volition.

Playing to be part of a community and playing for challenge were rated equally as reasons from gaming. This supports that social aspects of play are important to our sample. But that challenge is an equally motivating, core component of game play for these players much like non-disabled mainstream gamers [7,9] Further investigation would be needed to find what community means to these players and whether they are referring to gaming communities in general or, more specifically, communities based around play with disabilities. Some further comments by participants give some indication that this may be community of players with disabilities, examples include:

> "To try to help others to find ways a person with a disability can play and enjoy gaming"
> "To show other people that it's possible and build a community for other disabled gamers"
> "To advocate"
> "To Inspire"

This finding may be a result of the participants choice to belong to and support the work of a charity advocating and facilitating play for a community of players with disabilities. Nonetheless, this is further evidence that the gaming hobby revolves around, and serves to establish communities and this supports that gaming can be treated as a social activity.

It is important to note that this sample of disabled players is likely subject to selection bias: these are players who currently play digital games and could complete our survey. As this survey was conducted to gain an overview of the AbleGamers Player Panels community, there are items which were not covered initially that could form the basis of further work. This may include covering: what assistive technologies or accessibility options/software players feel that they do not have but would help them, a broader look at what gaming platforms may be used (e.g. older consoles such as Nintendo Wii), a deeper look at who they are playing with and what their online multiplayer experiences are like.

More importantly, even though there will likely always be a need to address the implementation lag of new technologies to provide accessible options [21], we see that commercial mainstream games are reaching a point in the research domain where there is the opportunity to move beyond simply providing access to games. There is the opportunity to explore what it means for players to have accessible player experiences within games, leveraging the existing wealth of knowledge from the player experience research community.

## 5     Conclusions

The demographic survey we conducted shows that our participants are much like samples of the wider population of players. They are playing mainstream games, they play online, and multiplayer games with both friends and strangers, they identify as 'gamers', and give substantial amounts of their free time to the hobby. Since previous research has focused on using games for therapeutic uses and rehabilitation, this work shows that, although this may motivate some people, players with disabilities are also playing for similar reasons as non-disabled players; for run, relaxation, challenge and community. Additionally, there may still be issues with control mechanisms for disabled players and mainstream games may not be entirely accessible, however despite this, there are still disabled players who do have access and do play popular mainstream games. Therefore, game designers and researchers can assume that people with disabilities want to play mainstream games with everyone else and will attempt to find a way to play. In terms of game design, since many of these players have reported using adaptations such as auditory alerts, key remapping, subtitles, alternative controllers, screen readers, and so on, this suggests that these minimal additions and modifications to games can accommodate for a substantial audience of disabled players. As such, it is becoming increasingly important for researchers and designers to consider not only the effectiveness of these adaptations but how these impact their overall APX of digital games and, consequently, their social experiences in playing games with others.

**Acknowledgments.** Thanks to the AbleGamers Charity for the collaboration on the Player Panels programme, and all the players who volunteered their time. This work is funded by EPSRC grant [EP/L015846/1] (IGGI) and the University of York Research Priming Fund.